\documentstyle[12pt]{article}
\newcommand{\eq}{\begin{equation}}
\newcommand{\en}{\end{equation}}
\newcommand{\eqa}{\begin{eqnarray}}
\newcommand{\ena}{\end{eqnarray}}
\newcommand{\eqs}{\begin{displaymath}}
\newcommand{\ens}{\end{displaymath}}
\newcommand{\eqas}{\begin{eqnarray*}}
\newcommand{\enas}{\end{eqnarray*}}

\advance\oddsidemargin by -\textwidth
\advance\oddsidemargin by -1in
\evensidemargin=\oddsidemargin
\begin{document}

$\mbox{ }$
\vspace{-3cm}
\begin{flushright}
\begin{tabular}{l}
{\bf KEK-TH-378 }\\
{\bf KEK preprint 93 }\\
December 1993
\end{tabular}
\end{flushright}

\baselineskip18pt
\vspace{1cm}
\begin{center}
\Large
{\baselineskip26pt \bf
String Field Theory \\
of $c\leq 1$ Noncritical Strings}
\end{center}
\vspace{1cm}
\begin{center}
\large
{\sc Nobuyuki Ishibashi and Hikaru Kawai}
\end{center}
\normalsize
\begin{center}
{\it KEK Theory Group, Tsukuba, Ibaraki 305, Japan}
\end{center}
\vspace{2cm}
\begin{center}
\normalsize
ABSTRACT
\end{center}
{\rightskip=2pc 
\leftskip=2pc 
\normalsize
We construct a string field Hamiltonian for a noncritical string theory
with the continuum limit of the Ising model or its generalization as the matter
theory on the worldsheet. It consists of only three string vertices
as in the case for $c=0$. We also discuss a general consistency condition
that should be satisfied by this kind of string field Hamiltonian.
\vglue 0.6cm}

\newpage
\section{Introduction}
\hspace{5mm}
String field theory is considered to be the
most promising formalism for giving a nonperturbative definition of string
theory
 \cite{SFT}.
In order to construct a string field theory, one should give a rule for
decomposing the worldsheets of the string into propagators and vertices.
This can be done by introducing a time coordinate
on the worldsheets. There are several ways of doing so and a string field
theory
corresponds to each of them. Recently, a new kind of time coordinate
, which can be naturally defined on dynamically
triangulated worldsheets,
was discovered\cite{KKMW}.
Using this time coordinate, a string field theory
for $c=0$ noncritical string field theory was constructed \cite{IK}. We showed
that the matrix model techniques \cite{DS}\cite{FKN},
which were used to discuss a nonperturbative aspect of the
noncritical string theory, can be easily deduced in this
string field theory formulation. Therefore, we expect that such a time
coordinate is also useful in constructing a string field theory for critical
strings.
In order to do so, we should know how to introduce matter field degrees of
freedom on the worldsheet in the string field theory.

In the present work, we will construct a string field theory for $c\leq 1$
noncritical strings. The matter conformal field theory will be represented by
the continuum limit of the Ising model or its generalization.
In our string field formalism, the matter degrees of freedom will be
taken into account
by introducing several kinds of string fields. The string field Hamiltonian
consists of only three string interaction terms. We will also discuss
general consistency conditions that should be satisfied by this kind of string
field theory.

\section{String Field Theory for $c=\frac{1}{2}$ String}
\hspace{5mm}
In this section we will construct a string field Hamiltonian for
$c=\frac{1}{2}$
string. The $c=\frac{1}{2}$ matter theory is represented by the continuum
imit of the Ising model.

Let us briefly recall the results of \cite{IK}, in which a string field
Hamiltonian for $c=0$ string theory was constructed. For the
$c=0$ string theory,
the string field depends on the length of the string, which is the only
diffeomorphism invariant quantity of a loop in this case. We defined the
creation (annihilation) operator, $\Psi^\dagger (l)$ ($\Psi (l)$),
 of a string with length $l$ by the following commutation relation:
\eq
\mbox{[} \Psi (l), \Psi^{\dagger }(l') \mbox{]}=\delta (l-l').
\label{comm}
\en
The string field Hamiltonian describes evolutions of a string
in the coordinate frame defined in \cite{KKMW}. It was written in terms of
$\Psi^\dagger (l)$ and $\Psi (l)$ as
\eqa
{\cal H}
&=&
\int_0^{\infty}dl_1\int_0^{\infty}dl_2
  \Psi^{\dagger}(l_1)\Psi^{\dagger}(l_2)\Psi (l_1+l_2)(l_1+l_2)
\nonumber
\\
& &
  +g\int_0^{\infty}dl_1\int_0^{\infty}dl_2
  \Psi^{\dagger}(l_1+l_2)\Psi (l_1)\Psi (l_2)l_1l_2
\nonumber
\\
& &
+\int _0^{\infty}dl\rho (l)\Psi (l).
\label{ham0}
\ena

We now consider the noncritical string theory described by
two dimensional gravity coupled to the Ising model. We will dynamically
triangulate the worldsheet and put the Ising spins on the vertices.
In this formulation,
the diffeomorphism invariant quantities of a loop are the configuration
of the spins on the loop as well as its length. Therefore
the string field should depend on the spin configuration on the string.
Here we will
restrict ourselves to the simplest configurations where all the spins on the
loop are the same, up or down. We denote the creation (annihilation) operator
for the strings with such a configuration as $\Psi^\dagger_\pm (l)$
($\Psi_\pm (l)$), where the subscript $\pm$ denotes the up or down spins on the
string. We will construct a string field Hamiltonian involving only these
fields. Since the usual matrix model technique treats only loops
with such spin
configurations, such a Hamiltonian is enough to reproduce the matrix model
results.

If we constructed
a straightforward generalization of the string field Hamiltonian in \cite{IK}
 using the time coordinate in \cite{KKMW},
we would need to take other spin configurations into account.
For example, let us consider a dynamically triangulated disk as the worldsheet.
Since Ising spins are
on the disk,  it is divided into the domains of up spins and down spins.
Suppose we
start with a string with all spins up. After the ``one-step deformation'' in
\cite{KKMW}, however,
down spins will appear,
if the string hits the domain walls. Therefore, in order to
make a string field theory by only two kinds of fields above, one should change
the definition of the time coordinate a bit. We can do so as follows.

We are going to modify the ``one-step deformation'' in \cite{KKMW} so that
only strings with all the spins aligned appear. Suppose we consider the
evolution of a string starting from, say, all spin up configuration.
Let us adopt the usual
definition of the ``one-step deformation'' until the string hits the domain
walls. When the string hits a domain wall, down spins appear in the string.
In this case
 we change the notion of the ``one-step deformation'' as follows.
Since the domain
wall divides the regions of up and down spins, along each side of the wall
only up or down spins appear. Therefore we consider that
the string has split into
two strings along the wall. The two strings are with all spins up and down
respectively (see Fig. 1). If the string hits many domain walls simultaneously,
we generalize the above definition and consider that
 the string splits into many with
all spins aligned.

With such a definition of the ``one-step deformation'', we can construct
a Hamiltonian describing the evolution. It should include the following
processes:
\begin{enumerate}
\item A string with all the spins up or down splits into the same kind of
strings.
\item A string with all the spins up or down splits into several with all spins
up and the others with all spins down.
\item A string disappears.
\end{enumerate}
The first and third processes are the ones we encountered in $c=0$ case
\cite{IK}. The second one comes from the modified
definition of the time evolution.
If one wants to consider worldsheets other than the disk, one should include
one more process:
\begin{description}
\item[$~~~$4.] Strings of the same kind merge.
\end{description}

The discrete Hamiltonian may consist of infinitely many terms of these
kinds.
In the continuum limit, however,
 only the terms which have the right scaling dimension
survive. It will take a formidable task to construct the discrete Hamiltonian
and take the continuum limit.
Here we rather conjecture the continuum Hamiltonian and show evidences for
its validity afterwards. The Hamiltonian we propose is
\eqa
{\cal H}
&=&
\int_0^{\infty}dl_1\int_0^{\infty}dl_2
  \Psi^{\dagger}_+(l_1)\Psi^{\dagger}_+(l_2)\Psi_+ (l_1+l_2)(l_1+l_2)
\nonumber
\\
& &
+\int_0^{\infty}dl_1\int_0^{\infty}dl_2
  \Psi^{\dagger}_+(l_1+l_2)\Psi^{\dagger}_-(l_2)\Psi_+ (l_1)l_1
\nonumber
\\
& &
  +g\int_0^{\infty}dl_1\int_0^{\infty}dl_2
  \Psi^{\dagger}_+(l_1+l_2)\Psi _+(l_1)\Psi _+(l_2)l_1l_2
\nonumber
\\
& &
+\mbox{[}~\Psi_+(\Psi^\dagger_+)~\longleftrightarrow ~\Psi_-(\Psi^\dagger_-)~
\mbox{]}.
\label{ham}
\ena
The first term and the third term are similar to the ones in
$c=0$ case. The second
term corresponds to the new kind of process: the process 2 above.
In such a process, as is clear from Fig. 1, a string with length
$l_1$ splits into two strings with lengths $l_1+l_2$ and $l_2$, where
$l_2$ is the length of the domain wall which the string hits.

If one expresses the
dimension of length $l$ by $L$, one has
$\mbox{[}\Psi^\dagger \mbox{]}=L^{-\frac{7}{3}},~
\mbox{[}\Psi \mbox{]}=L^{\frac{4}{3}}$ and
$\mbox{[}{\cal H}\mbox{]}=L^{-\frac{1}{3}}$.
As in the case for $c=0$, the dimension of $\Psi^\dagger $ should coincide with
that of the disk amplitude, which is $L^{-\frac{7}{3}}$ \cite{KPZ} for
$c=\frac{1}{2}$.
Therefore
the dimension of the time coordinate corresponding to this Hamiltonian is
$L^{\frac{1}{3}}$. Thus, admitting that the Hamiltonian in eq.(\ref{ham}) is
the continuum limit of the discrete Hamiltonian describing the processes above,
we can consider that the continuum limit of the above ``one-step deformation''
defines a distance whose dimension is $L^{\frac{1}{3}}$ on the worldsheet of
$c=\frac{1}{2}$ string theory.
The string coupling constant $g$ has the dimension
$L^{-\frac{14}{3}}$, which is consistent with the matrix model result.
Contrary to the $c=0$ case, however,
this Hamiltonian does not include a tadpole term.
We expect that the tadpole term can be written in terms of the derivatives
of $\delta (l)$ and involves only integer powers of the
cosmological constant $t$. Hence, only strings with vanishing
length can disappear
and the worldsheet in such a process has zero area. However there is no such
term with the right scaling dimension.

In the following, we will check if
this Hamiltonian really describes $c=\frac{1}{2}$ noncritical string.
In the rest of this section,
we will show that we can reproduce at least
the disk amplitudes of $c=\frac{1}{2}$
string theory from this Hamiltonian. In the next section,
we will discuss a consistency
condition that should be satisfied by this kind of string field Hamiltonian.
It will
easy to see that the Hamiltonian in eq.(\ref{ham}) satisfies such a
condition.

The disk amplitude can be expressed
by the Hamiltonian as in the case for $c=0$.
Let $f_+ (l)$ ($f_-(l)$) be the disk partition function with the boundary whose
length
is $l$ and whose spins are all up ( down).
If one defines the bra and ket
vacuum $<0|$ and $|0>$ by
\eqs
\Psi_\pm (l)|0>=<0|\Psi^{\dagger}_\pm (l)=0,
\ens
the disk amplitudes $f_\pm (l)$ can be expressed as
\eq
f_\pm (l) =\lim_{D\rightarrow \infty}
<0|e^{-D{\cal H}}\Psi^{\dagger}_\pm (l)|0>\mid_{g=0}.
\label{disk}
\en
We can calculate $f_\pm (l)$ by deriving the Schwinger-Dyson (S-D) equation
from this expression and then solving it. As in \cite{IK}, the S-D equation
is given by
\eq
\lim_{D\rightarrow \infty}\frac{\partial}{\partial D}
      <0|e^{-D{\cal H}_{disk}}\Psi^{\dagger}_\pm (l)|0>\mid_{g=0}=0.
\label{partD}
\en
Using the factorization property of the amplitudes for $g=0$, we obtain
\eq
l\int_0^{l}dl'f_\pm (l')f_\pm (l-l')
+l\int_0^{\infty}dl'f_\pm (l+l')f_\mp (l')=0.
\label{SDd}
\en

We now solve this equation. The $Z_2$ symmetry of the Ising model
implies $f_+(l)=f_-(l)\equiv f_{\frac{1}{2}}(l)$.
Therefore, Laplace transforming it, eq.(\ref{SDd}) becomes
\eq
\partial_\zeta \mbox{[}
(\tilde{f}_{\frac{1}{2}}(\zeta ))^2+
\frac{1}{2\pi i}\int_C\frac{d\zeta ^\prime }{\zeta -\zeta^\prime}
\tilde{f}_{\frac{1}{2}}(\zeta^\prime )\tilde{f}_{\frac{1}{2}}(-\zeta^\prime )
\mbox{]}=0.
\label{kostov}
\en
Here $\tilde{f}_{\frac{1}{2}}(\zeta )
=\int_0^\infty dle^{-\zeta l}f_{\frac{1}{2}}(l)$ and $C$ is the
contour in the complex $\zeta^\prime$ plane depicted in Fig.2.

In general, the integral in eq.(\ref{kostov})
is not well-defined because of the divergence of the integrand
as $\zeta^\prime \rightarrow \infty$.
In order to deal with such an ultraviolet divergence, one needs a
consistent regularization. Actually such a regularization is known.
Indeed,
eq.(\ref{kostov}) is exactly the kind of equation that Kostov has encountered
in
his analyses of $ADE$ lattice models on a dynamically triangulated
surface\cite{kostov}. Therefore if one considers that the Hamiltonian in
eq.(\ref{ham}) describes the continuum limit of such a discrete model, the
method
to regularize the ultraviolet divergence is apparent.

For example, one can solve eq.(\ref{kostov}) by
symmetrizing it with respect to the reflection
$\zeta \leftrightarrow -\zeta $:
\eqs
\partial_\zeta \mbox{[}
(\tilde{f}_{\frac{1}{2}}(\zeta ))^2
+\tilde{f}_{\frac{1}{2}}(\zeta )\tilde{f}_{\frac{1}{2}}(-\zeta )
+(\tilde{f}_{\frac{1}{2}}(-\zeta ))^2
\mbox{]}=0.
\ens
In doing so, an equality
\eq
\int_0^\infty dle^{-\zeta l}\int_0^{\infty}dl'
f_{\frac{1}{2}}(l+l')f_{\frac{1}{2}}(l')
+(~\zeta \leftrightarrow -\zeta ~)
=\tilde{f}_{\frac{1}{2}}(\zeta )\tilde{f}_{\frac{1}{2}}(-\zeta ),
\label{equa}
\en
is used, which can be derived when one regularizes it as in \cite{kostov}.
The cosmological constant $t$ with the dimension $\mbox{[}t\mbox{]}=L^2$ is
then introduced as the integration constant:
\eq
(\tilde{f}_{\frac{1}{2}}(\zeta ))^2
+\tilde{f}_{\frac{1}{2}}(\zeta )\tilde{f}_{\frac{1}{2}}(-\zeta )
+(\tilde{f}_{\frac{1}{2}}(-\zeta ))^2
=3t^{\frac{4}{3}}.
\label{ez}
\en
Eq.(\ref{ez}) can be solved \cite{kostov}\cite{ez} to give
\eq
\tilde{f}_{\frac{1}{2}}(\zeta )=
(\zeta +\sqrt{\zeta^2 -t})^{\frac{4}{3}}+
(\zeta -\sqrt{\zeta^2 -t})^{\frac{4}{3}},
\label{disso}
\en
which coincides with
the disk amplitude for the $c=\frac{1}{2}$ noncritical string
theory\cite{MSS}.

One can proceed further and derive the S-D equation for more general
amplitudes. Let us define the generating functional $Z(J_+,J_-)$ of
loop operators by
\eq
Z(J_+,J_-)=
\lim_{D\rightarrow \infty}
<0|e^{-D{\cal H}}e^{\int dl\mbox{[}
                    J_+(l)\Psi^{\dagger }_+(l)+
                    J_-(l)\Psi^{\dagger }_-(l)
                    \mbox{]}}|0>.
\en
The S-D equation for this quantity can be derived as in eq.(\ref{partD}) and
one obtains
\eq
\int_0^\infty dll\mbox{[}J_+(l)T_+(l)+J_-(l)T_-(l)\mbox{]}Z(J_+,J_-)=0,
\label{eqZ}
\en
where
\eqa
T_\pm (l)
&=&
\int_0^ldl'
\frac{\delta^2}{\delta J_\pm (l')\delta J_\pm (l-l')}
\nonumber
\\
& &
+\int_0^\infty dl'
\frac{\delta^2}{\delta J_\pm (l+l')\delta J_\mp (l')}
\nonumber
\\
& &
+g\int_0^{\infty }dl'J_\pm (l')l'
\frac{\delta }{\delta J_\pm (l+l')}.
\label{Tpm}
\ena
Therefore if the equation
\eq
lT_\pm (l)Z(J_+,J_-)=0
\label{sdeq}
\en
has a solution satisfying the appropriate
boundary conditions, it is a solution of eq.(\ref{eqZ}).

By expanding
eq.(\ref{sdeq}) in terms of $g$, we can obtain equations for amplitudes with
various topologies. By solving these, one can in principle calculate the
string amplitudes with any number of handles and local operator insertions.
For example, it is possible to show that the disk amplitudes
with one local operator insertion of the $c=\frac{1}{2}$ string \cite{MSS}
are the solutions of such equations.  We expect that the same thing happens
for amplitudes with more local operator insertions and more handles.
Unfortunately it seems difficult to
prove that the Hamiltonian describes the $c=\frac{1}{2}$ string by showing
that eq.(\ref{sdeq}) is equivalent to the $W$ constraints \cite{FKN},
contrary to the $c=0$ case. As is clear from the disk example in eq.(\ref{ez}),
the S-D equations in this formulation do not give algebraic equations
after the Laplace transformation. One should employ some transcendental
techniques to solve it. Hence it is difficult to see the direct relation
between our S-D equation and the $W$ constraints, which are algebraic
equations of loop amplitudes.

\section{The Consistency Condition}
\hspace{5mm}
In the preceeding section,
 we proposed a string field Hamiltonian eq.(\ref{ham}) for $c=\frac{1}{2}$
noncritical string and checked that it reproduced the disk amplitudes of
the matrix model. In this section, we will give another evidence for the
validity of this Hamiltonian. We will propose a consistency condition
that should be satisfied by
this kind of string field Hamiltonian. This condition plays a
similar role as the modular invariance did in the string model building.
We expect that it
restricts the form of string field couplings severely.
Since the Hamiltonian in eq.(\ref{ham}) will be shown to
satisfy the condition, we consider it
as the right Hamiltonian for $c=\frac{1}{2}$ string theory.

Let us first derive the consistency condition for
$c=0$ string theory and show that our string field Hamiltonian satisfies it.
We will consider the worldsheet with the topology of the cylinder. Suppose
we have
a cylinder in which the
minimum geodesic distance of the two boundaries of it is $D$ and the lengths
of the boundaries are $l_1$ and $l_2$.
We can calculate the amplitude corresponding to such a configuration in the
formalism of \cite{KKMW}\cite{IK}. Let us introduce the time coordinate on the
worldsheet as in these references and follow the evolutions of the strings.
Starting from the two boundaries, they keep splitting and disappearing until
the time coordinate reaches $\frac{D}{2}$. At the time $\frac{D}{2}$,
one string coming from one boundary and another string from the other boundary
merge. After merging, the resultant string keeps splitting and disappearing to
form a disk (Fig. 3). Therefore we have the following expression for such an
amplitude:
\eq
\int_0^\infty dl_1^\prime \int_0^\infty dl_2^\prime
f(l_1^\prime +l_2^\prime )l_1^\prime l_2^\prime
<0|\Psi (l_1^\prime )e^{-\frac{D}{2}H}\Psi^\dagger (l_1)|0>
<0|\Psi (l_2^\prime )e^{-\frac{D}{2}H}\Psi^\dagger (l_2)|0>.
\label{D/2}
\en
Here $f(l)$ is the disk amplitude for $c=0$ string and $H$ is the Hamiltonian
for the {\it inclusive} process\cite{KKMW}\cite{IK}:
\eqs
H=2\int_0^\infty dl_1\int_0^\infty dl_2f(l_1)
\Psi^\dagger (l_2)\Psi (l_1+l_2)(l_1+l_2).
\ens

However the above is not the unique way of calculating such an amplitude.
We can arrange the time coordinate on the worldsheet so that the merging
occurs at a point from which the minimum geodesic distance to one boundary is
$D_1$ and to the other is $D_2$ with $D_1+D_2=D$ (Fig. 3).
Such a time
coordinate is possible, if one does not make the two boundaries start evolving
simultaneously. In such a coordinate system,
the amplitude can be calculated in a similar way:
\eq
\int_0^\infty dl_1^\prime \int_0^\infty dl_2^\prime
f(l_1^\prime +l_2^\prime )l_1^\prime l_2^\prime
<0|\Psi (l_1^\prime )e^{-D_1H}\Psi^\dagger (l_1)|0>
<0|\Psi (l_2^\prime )e^{-D_2H}\Psi^\dagger (l_2)|0>.
\label{D1D2}
\en

Since we are calculating the same amplitude, the results (eqs.(\ref{D/2}) and
(\ref{D1D2})) should coincide as long as $D_1+D_2=D$. Namely
in order for the theory to be general coordinate invariant, the amplitudes
should not change when one changes the coordinate system.
Such a condition implies the following equality for
the disk amplitude $f(l)$:
\eqa
& &
l_1l_2\mbox{[}
\int_0^{l_1}dl_1^\prime
\int_0^{\infty}dl_2^\prime
f(l_1^\prime )f(l_2^\prime )(l_1-l_1^\prime )
\delta (l_1+l_2-l_1^\prime -l_2^\prime )
\nonumber
\\
& &
-
\int_{l_1}^\infty dl_1^\prime
\int_0^{\infty}dl_2^\prime
f(l_1^\prime )f(l_2^\prime )(l_1-l_1^\prime )
\delta (l_1+l_2-l_1^\prime -l_2^\prime )
\mbox{]}
=0.
\label{ameq}
\ena
This is the consistency condition that should be satisfied
by the disk amplitude.
Each term on the left hand side of this equation
is made from a combination of two kinds of the three string vertices in the
Hamiltonian eq.(\ref{ham0}). In Fig. 4, we schematically present such
combinations. The left hand side of Fig. 4 corresponds to that
of eq.(\ref{ameq}).

The equality in eq.(\ref{ameq}) can be proved by reducing it to the disk S-D
equation.
Symmetrizing in terms of $l_1^\prime $ and $l_2^\prime $,
the left hand side of eq.(\ref{ameq}) eventually becomes
\eq
(l_1-l_2)l_1l_2\int_0^{l_1+l_2}dl^\prime f(l^\prime )f(l_1+l_2-l^\prime ).
\label{l1+l2}
\en
Eq.(\ref{l1+l2}) can be
illustrated as the right hand side of Fig. 4. The factor $l_1l_2$ is
accounted for by the merging vertex. Therefore the
transformation from eq.(\ref{ameq}) to eq.(\ref{l1+l2}) can be considered
as the duality-like transformation depicted in Fig. 4.

In this form the consistency condition can be related to
the disk S-D equation
\eqs
l\int_0^{l}dl^\prime f(l^\prime )f(l-l^\prime )
-
3\delta ^{\prime \prime }(l)+\frac{3}{4}t\delta(l)=0,
\ens
with the argument $l$ replaced by $l_1+l_2$.
Substituting the disk S-D equation, we can show that the consistency condition
can be proved, if
\eq
l_1l_2\frac{l_1-l_2}{l_1+l_2}\mbox{[}
3\delta ^{\prime \prime }(l_1+l_2)-\frac{3}{4}t\delta(l_1+l_2)
\mbox{]}
=0.
\label{subt}
\en
Eq.(\ref{subt}) is subtle because
both $l_1$ and $l_2$ vanish at
the support, $l_1+l_2=0$, of the delta functions, provided $l_1,l_2\geq 0$.
One way to deal with it is to
Laplace transform it. Using
\eq
\int_0^\infty dl_1\int_0^\infty dl_2
e^{-\zeta_1l_1-\zeta_2l_2}h(l_1+l_2)
=
-\frac{\tilde{h}(\zeta_1)-\tilde{h}(\zeta_2)}{\zeta_1-\zeta_2},
\en
where $\tilde{h}(\zeta )=\int_0^\infty dle^{-\zeta l}h(l)$, one can show
that the Laplace transformation of the left hand side of eq.(\ref{subt})
vanishes.

Thus our $c=0$ string field Hamiltonian satisfies the above proposed
consistency condition for the cylindrical amplitude.
The consistency condition was proved by
reducing it to the disk S-D equation via the duality-like transformation
in Fig. 4. It is rather miraculous that such a transformation exists. However,
if one rewrites the consistency condition in the following way, the meaning of
the duality-like transformation becomes clearer.
The S-D equation for the generating functional $Z(J)$ of the
$c=0$ string loop amplitudes can be written as
\eq
\mbox{[}lT(l)+\rho (l)\mbox{]}Z(J)=0,
\label{SDg}
\en
where
\eqa
T(l)
&=&
\int_0^ldl'
\frac{\delta^2}{\delta J(l')\delta J(l-l')}
\nonumber
\\
& &
+g\int_0^{\infty }dl'J(l')l'
\frac{\delta }{\delta J(l+l')},
\label{tl}
\ena
and
\eq
\rho (l)
=
\delta ^{\prime \prime }(l)-\frac{3}{4}t\delta (l).
\en
Eq.(\ref{SDg}) implies
\eq
\mbox{[}l_1T(l_1)+\rho (l_1),l_2T(l_2)+\rho (l_2)\mbox{]}Z(J)=0.
\label{intc}
\en
Actually the order $g$ part of the left hand side of
eq.(\ref{intc}) coincides with the left hand side of
eq.(\ref{ameq}). Therefore the consistency condition is included as the
order $g$ coefficient in
eq.(\ref{intc}), which is deduced from the S-D equation eq.(\ref{SDg}).
The higher order terms of
eq.(\ref{lvir}) may be interpreted as the higher genus generalization of
the cylindrical consistency
condition.

The functional differential equation eq.(\ref{SDg}) is integrable.
Indeed, it is easy to calculate the
commutation relation of the operator $lT(l)+\rho (l)$:
\eq
\mbox{[}l_1T(l_1)+\rho (l_1),l_2T(l_2)+\rho (l_2)\mbox{]}
=gl_1l_2\frac{l_1-l_2}{l_1+l_2}
\mbox{[}(l_1+l_2)T(l_1+l_2)+\rho (l_1+l_2)\mbox{]}
-gl_1l_2\frac{l_1-l_2}{l_1+l_2}\rho (l_1+l_2).
\label{lvir}
\en
The last term on the right hand side
is precisely the left hand side of eq.(\ref{subt}) and it vanishes.
Hence eq.(\ref{intc}) can be proved by using eq.(\ref{lvir}) and reducing it
to the S-D equation. At the order $g$, eq.(\ref{lvir}) is exactly the
duality-like transformation Fig. 4. Thus the duality-like transformation is
equivalent to the integrability condition of the S-D equation.

Since eq.(\ref{SDg}) is equivalent to the Virasoro constraints \cite{FKN},
it is natural to expect the algebra in eq.(\ref{lvir}) is actually
the Virasoro algebra. Indeed, if one expands the Laplace transform
$\tilde{T}(\zeta )=\int_0^\infty dle^{-\zeta l}T(l)$ as
\eq
\tilde{T}(\zeta )=
\sum_{n=-1}^\infty L_n\zeta^{-n-2},
\en
the algebra
\eq
\mbox{[}T(l_1),T(l_2)\mbox{]}=g(l_1-l_2)T(l_1+l_2)
\label{convi}
\en
coincides with
\eq
\mbox{[}L_n,L_m\mbox{]}=(n-m)L_{n+m}\;(m,n\geq -1).
\label{disvi}
\en

It is intriguing to observe that the algebra in eq.(\ref{convi}) looks like
the ``continuum limit'' of the Virasoro algebra eq.(\ref{disvi}) in the
following sense:\footnote{It may be possible to show that such a ``continuum
limit'' is indeed the continuum limit of the Virasoro algebra discovered in
\cite{IM}.
A similar observation is made by Y.Matsuo\cite{mats}.}
\eqa
n(\geq -1)
&\longrightarrow &
l(\geq 0),
\nonumber
\\
L_n
&\longrightarrow &
\frac{1}{g}T(l),
\nonumber
\\
\mbox{[}L_n,L_m\mbox{]}=(n-m)L_{n+m}
&\longrightarrow &
\mbox{[}\frac{1}{g}T(l_1),\frac{1}{g}T(l_2)\mbox{]}=
(l_1-l_2)\frac{1}{g}T(l_1+l_2).
\ena
That is to say,
the structure constant $l_1-l_2$ of the algebra of $T(l)$ is exactly the
``continuum limit'' of the structure constant $n-m$ of the Virasoro algebra.
Moreover the form of $T(l)$ in eq.(\ref{tl}) is the ``continuum limit''
of the Virasoro operator made from the bosonic oscillator:
\eqa
\alpha_n\;(n>0)
&\longrightarrow&
\frac{\delta}{\sqrt{g}\delta J(l)},
\nonumber
\\
\alpha_n\;(n<0)
&\longrightarrow&
\sqrt{g}lJ(l),
\nonumber
\\
\mbox{[}\alpha_n,\alpha_{-m}\mbox{]}=n\delta_{n,m}
&\longrightarrow&
\mbox{[}\frac{\delta}{\sqrt{g}\delta J(l)},\sqrt{g}l'J(l')\mbox{]}
=l\delta (l-l'),
\nonumber
\\
L_n=\frac{1}{2}\sum_m\alpha_{n-m}\alpha_m
&\longrightarrow&
T(l)
=
\int_0^ldl'
\frac{\delta^2}{\delta J(l')\delta J(l-l')}
+g\int_0^{\infty }dl'J(l')l'
\frac{\delta }{\delta J(l+l')}.
\nonumber
\\
& &
\ena
This kind of observation is useful in supersymmetrizing our formalism
\cite{IKN}.

Thus far
we have proposed the consistency condition of the string field Hamiltonian
for $c=0$ string and proved it by using the duality-like transformation in
Fig. 4. We showed that such a transformation is included in the integrability
condition of the S-D equation as the order $g$ coefficient.
We now examine whether the Hamiltonian eq.(\ref{ham}) for $c=\frac{1}{2}$
string
satisfies a similar condition.
The Hamiltonian is related to the modification of the one-step deformation
defined in section 2. In order to consider the consistency condition for the
cylindrical amplitude in this case, we should use the distance which can be
obtained as the continuum limit of this modified one-step deformation.
We start from a cylinder in which the minimum of such distance between the
two boundaries is $D$. Then we can proceed as in $c=0$ case and obtain
a consistency condition.

On the other hand, the integrability condition of S-D equation in this case
means the
closure of the commutation relations of the operators $T_\pm (l)$ in
eq.(\ref{Tpm}). One can show
\eqa
\mbox{[}T_\pm (l_1),T_\pm (l_2)\mbox{]}
&=&
g(l_1-l_2)T_\pm (l_1+l_2),
\nonumber
\\
\mbox{[}T_\pm (l_1),T_\mp (l_2)\mbox{]}
&=&
0.
\label{multv}
\ena
Namely, we have two decoupled Virasoro constraints.
Again the order $g$ part of the left hand sides of the
above equations exactly reproduce the form of the
consistency condition for the cylindrical
amplitude. Eq.(\ref{multv}) implies a generalization of the duality-like
transformation in Fig. 4. It enables us to prove that
the Hamiltonian in eq.(\ref{ham}) satisfies the
consistency condition.

\section{$c\leq 1$ Strings}
\hspace{5mm}
It is straightforward to generalize the Hamiltonian in eq.(\ref{ham}) to the
case where the matter theory on the worldsheet is a $c\leq 1$ unitary conformal
field theory.
Let us dynamically triangulate the worldsheet.
The matter theory is introduced by putting height variables on the vertices of
the triangulated surface.
Suppose that
the height variables take their values on the nodes of an $ADE$ or
$\hat{A}\hat{D}\hat{E}$ type
Dynkin diagram and obey the following rules.
\begin{enumerate}
\item The heights for the neighbouring sites should be the same or linked
on the Dynkin diagram.
\item At least two of the three heights assigned to a triangle should be the
same.
\end{enumerate}
The Ising model is the simplest ($A_2$) case.
We will consider matter theories realized by such lattice models.
The models considered here is similar to the ones discussed in \cite{pas} and
we expect that the continuum limits of them cover the $c\leq 1$ unitary
conformal field theories.

Now we will construct the string field Hamiltonian corresponding to the
matter theory. As in the Ising case,
we will deal with the strings on which the heights
take the same value. Hence we should prepare the string fields
$\Psi_i$, where the subscript $i$ denotes the height. The worldsheets
are divided into domains in which the height variables take the same value.
As long as a string travels in a domain, all the heights on the string are
the same. Different heights come in when the string hits the domain walls.
Let us adopt the same definition of the one-step deformation
as in the Ising model for such a case.
Assuming that there exists a continuum limit of this system, we propose
the following Hamiltonian describing the time evolution of the string
in the coordinate system, which is an obvious generalization of the Hamiltonian
in eq.(\ref{ham}):
\eqa
{\cal H}
&=&
\sum_i\int_0^{\infty}dl_1\int_0^{\infty}dl_2
  \Psi^{\dagger}_i(l_1)\Psi^{\dagger}_i(l_2)\Psi_i (l_1+l_2)(l_1+l_2)
\nonumber
\\
& &
+\sum_{i,j}C_{ij}\int_0^{\infty}dl_1\int_0^{\infty}dl_2
  \Psi^{\dagger}_i(l_1+l_2)\Psi^{\dagger}_j(l_2)\Psi_i  (l_1)l_1
\nonumber
\\
& &
  +g\sum_i\int_0^{\infty}dl_1\int_0^{\infty}dl_2
  \Psi^{\dagger}_i(l_1+l_2)\Psi _i(l_1)\Psi _i(l_2)l_1l_2.
\label{dham}
\ena
Here, $C_{ij}$ is the connectivity matrix, where $C_{ij}=1$ when the heights
$i$ and $j$ are linked on the Dynkin diagram and it vanishes otherwise.
This is a straightforward generalization of the Ising Hamiltonian in
eq.(\ref{ham}). In the rest of this section, we will argue that this
Hamiltonian
describes a $c\leq 1$ string theory.

It is easy to make a S-D equation for the disk amplitudes from this
Hamiltonian. Let $f_i(l)$ be the disk partition function in which the height
variables on the boundary are $i$. $f_i(l)$ can be expressed as
$f_i(l)=\lim_{D\rightarrow \infty}
<0|e^{-D{\cal H}}\Psi^{\dagger}_i(l)|0>\mid_{g=0}$ and the S-D equation
for $f_i$ becomes
\eq
l\int_0^{l}dl'f_i(l')f_i(l-l')
+l\sum_jC_{ij}\int_0^{\infty}dl'f_i(l+l')f_j(l')=0.
\label{dSDd}
\en
In the $ADE$ case, it has the solution of the form
\eq
\tilde{f}_i(\zeta )
=
v_i\mbox{[}
(\zeta +\sqrt{\zeta^2 -t})^\alpha +
(\zeta -\sqrt{\zeta^2 -t})^\alpha
\mbox{]},
\label{solf}
\en
where $\alpha$ and $v_i$ satisfy
\eq
\sum_j C_{ij}v_j=-2\cos (\pi \alpha )v_i.
\label{alv}
\en
Let us assume that our theory is unitary and $v_i>0$.
Eq.(\ref{alv}) has a solution
$\alpha =\frac{h+1}{h}$ with $v_i>0$, where $h$ is the Coxeter number of the
Dynkin diagram\cite{pas}.
This is exactly what we expect for the disk amplitude for
the unitary conformal field theory with $c=1-\frac{6}{h(h+1)}$
coupled to quantum gravity
\cite{MSS}. In the $\hat{A}\hat{D}\hat{E}$ case, the disk amplitudes of the
form
\eq
\tilde{f}_i(\zeta )
=
v_i\mbox{[}
(\zeta +\sqrt{\zeta^2 -t})\ln \mbox{[}\zeta +\sqrt{\zeta^2 -t}\mbox{]}+
(\zeta -\sqrt{\zeta^2 -t})\ln \mbox{[}\zeta -\sqrt{\zeta^2 -t}\mbox{]}
\mbox{]},
\en
satisfy the loop equation with $v_i>0$ and
\eq
\sum_j C_{ij}v_j=2v_i.
\en
This also coincides with the disk amplitudes of $c=1$ string theory.

It is easy to check that the consistency condition holds for
the Hamiltonian eq.(\ref{dham}). The integrability condition yields
decoupled Virasoro algebras again.
Therefore our Hamiltonian may be considered to
describe the $c\leq 1$ string theory.

One may well take the connectivity matrix $C_{ij}$ of an arbitrary diagram and
construct a Hamiltonian as in eq.(\ref{dham}). Assuming the solution of
the form in eq.(\ref{solf}),  the disk S-D equation is reduced to
eq.(\ref{alv}). Only $ADE$ and $\hat{A}\hat{D}\hat{E}$ Dynkin diagrams are
the ones in which eq.(\ref{alv}) has a solution with real $\alpha$ and positive
$v_i$. Since $ADE$ and $\hat{A}\hat{D}\hat{E}$ lattice models realize the
unitary conformal field theories with $c\leq 1$, this fact can be considered
as the manifestation of $c=1$ barrier in this formalism.

\section{Conclusions}
\hspace{5mm}
In this article we have constructed string field Hamiltonians for
$c\leq 1$ string
theories. In contrast with the $c=0$ case, we had to change the definition of
the time coordinate so that only the string fields with all the heights aligned
can appear. It seems that the $c=1$ barrier is present in this kind of
formulation. In order to generalize our string field theory to the critical
string case, a formulation which deals with the whole matter
configurations on the
string may be necessary. Work in this direction is in progress \cite{FIKN}.

Moreover,
in the present formulation, we restrict ourselves to the string theory with a
unitary matter conformal field theory on the worldsheet.
It may be possible to obtain nonunitary
matter theories as the multicritical phases of the unitary theory\cite{kaz}.
In a recent
work\cite{kle}, it was shown that the transfer matrix formalism in \cite{KKMW}
can be extended to the multicritical phases.

We have also discussed the consistency conditions that should be satisfied by
this kind of string field Hamiltonian. We have shown that
such conditions are related to the integrability condition for the S-D
equation.
The algebraic structure appearing in such an integrability condition
 may be a clue to construction of more general string field theories of
this type.

\section*{Acknowledgements}
We would like to thank N.Kawamoto, Y.Matsuo, T.Mogami, Y.Okamoto, Y.Watabiki,
Y.Yamada and T.Yukawa for useful discussions and comments.

\newpage
\section*{Figure Captions}
\begin{description}
\item{Fig. 1} One-step deformation when a string hits a domain wall.
When a string with all the spins up hits a domain wall, we
consider that it splits into two strings with all spins up and all spins down.
If the length of the incident string is $l_1$ and that of the domain wall is
$l_2$, a string with length $l_1+l_2$ and all spins up and another with
length $l_2$ and all spins down are generated.
\item{Fig. 2} The integration contour $C$ in the complex $\zeta '$ plane.
$C$ should
be taken so that it is on the left of $\zeta$ and goes between the two cuts of
the
integrand, $\zeta '> \sqrt{t}$ and $\zeta '< -\sqrt{t}$ on the real axis.
\item{Fig. 3} Two ways of computing the cylinder in which the minimum geodesic
distance between the two boundaries is $D$.
\item{Fig. 4} The duality-like relation.
\end{description}

\begin{thebibliography}{99}
\bibitem{SFT}
M.Kaku and K.Kikkawa, Phys. Rev. {\bf D10}(1974)1110;1823;\\
W.Siegel, Phys. Lett. {\bf B151}(1985)391;396;\\
E.Witten, Nucl. Phys. {\bf B268}(1986)253;\\
H.Hata, K.Itoh, T.Kugo, H.Kunitomo and K.Ogawa,
Phys. Lett. {\bf B172}(1986)186;195;\\
A.Neveu and P.West, Phys. Lett. {\bf B168}(1986)192.

\bibitem{KKMW}H.Kawai, N.Kawamoto, T.Mogami and Y.Watabiki, Phys. Lett.
{\bf B306}(1993)19.

\bibitem{IK}N.Ishibashi and H.Kawai, Phys. Lett. {\bf B314}(1993)190.

\bibitem{DS}E.Br\'{e}zin and V.Kazakov, Phys. Lett. {\bf B236}(1990)144;\\
M.Douglas and S.Shenker, Nucl. Phys. {\bf B335}(1990)635;\\
D.Gross and A.Migdal, Phys. Rev. Lett. {\bf 64}(1990)127;
Nucl. Phys. {\bf B340}(1990)333.

\bibitem{FKN}M.Fukuma, H.Kawai and R.Nakayama, Int. J. Mod. Phys.
{\bf A6}(1991)1385; \\
R.Dijkgraaf, E.Verlinde and H.Verlinde, Nucl. Phys. {\bf B348}(1991)435.

\bibitem{KPZ}V.G.Knizhnik, A.M.Polyakov and A.B.Zamolodchikov, Mod. Phys. Lett.
{\bf A3}(1988)819; \\
F.David, Mod. Phys. Lett. {\bf A3}(1988)1651; \\
J.Distler and H.Kawai, Nucl. Phys. {\bf B321}(1989)509.

\bibitem{kostov}I.K.Kostov, Nucl. Phys. {\bf B326}(1989)583;
 Phys. Lett. {\bf B266}(1991)42; Nucl. Phys. {\bf B376}(1992)539.

\bibitem{ez}B.Eynard and J.Zinn-Justin, Nucl. Phys. {\bf B386}(1992)558.

\bibitem{MSS}G.Moore, N.Seiberg and M.Staudacher, Nucl. Phys. {\bf B362}(1991)
665.

\bibitem{IM}H.Itoyama and Y.Matsuo, Phys. Lett. {\bf B255}(1991)202.

\bibitem{mats}Y.Matsuo, private communication.

\bibitem{IKN}N.Ishibashi, H.Kawai and J.Nishimura, in preparation.

\bibitem{pas}V.Pasquier, Nucl. Phys. {\bf B285}(1987)162;\\
P.Ginsparg, Nucl. Phys. {\bf B295}(1988)153.

\bibitem{FIKN}M.Fukuma, N.Ishibashi, H.Kawai and M.Ninomiya, in preparation.

\bibitem{kaz}V.A.Kazakov, Mod. Phys. Lett. {\bf A4}(1989)2125;\\
I.K.Kostov and M.Staudacher, Nucl. Phys. {\bf B384}(1992)459.

\bibitem{kle}S.S.Gubser and I.R.Klebanov, Preprint PUPT-1422,
hep-th@xxx.lanl.gov - 9310098;\\
A.Tsuchiya, unpublished.

\end{thebibliography}
\end{document}